# Adversarial Deep Learning for Simultaneous Segmentation of Ventricular and White Matter Hyperintensities in Clinical MRI

**Mahdi Bashiri Bawil**

Orcid: https://orcid.org/0009-0002-2029-3245 ; Email: m_bashiri99@sut.ac.ir

Affiliation: Biomedical Engineering Faculty, Sahand University of Technology, Tabriz, Iran

**Mousa Shamsi**[1]

Orcid: https://orcid.org/0000-0003-4670-0531 ; Email: shamsi@sut.ac.ir

Affiliation: Biomedical Engineering Faculty, Sahand University of Technology, Tabriz, Iran

**Abolhassan Shakeri Bavil**

Orcid: https://orcid.org/0000-0001-9397-0484 ; Email: shakeribavil@yahoo.com

Affiliation: Radiology Department, Tabriz University of Medical Sciences, Tabriz, Iran

Corresponding author: **Mousa Shamsi,** shamsi@sut.ac.ir

[1] Corresponding author

## Abstract

**Purpose**: Multiple sclerosis (MS) diagnosis requires accurate assessment of white matter hyperintensities (WMH) and ventricular changes on brain MRI. Current methods treat these structures independently, struggle to differentiate normal from pathological hyperintensities, and perform poorly on anisotropic clinical data. We present a deep learning framework that simultaneously segments ventricles and WMH while distinguishing normal periventricular hyperintensities from pathological MS lesions.

**Methods**: We developed a 2D pix2pix architecture trained on FLAIR scans from 300 MS patients combined with the MSSEG2016 benchmark (15 patients). Five architectural variants were compared through systematic ablation using 5-fold cross-validation with patient-level stratification, progressively integrating adversarial training, attention-weighted discrimination, and adaptive hybrid loss. Performance was assessed against six established methods using Dice coefficient, Hausdorff distance, precision, and recall.

**Results**: The final architecture (V5) achieved mean Dice 0.852±0.004 and HD95 4.87±0.13mm across all classes. Per-class performance: ventricles (Dice 0.907±0.002, HD95 3.00±0.51mm), abnormal WMH (Dice 0.825±0.009, HD95 4.51±0.32mm), normal WMH (Dice 0.677±0.007). V5 outperformed all baselines on local data for both ventricle and WMH segmentation. Ablation analysis confirmed adversarial training provided the largest single gain (+0.109 Dice). End-to-end processing required ~4 seconds per case—up to 36× faster than baseline methods.

**Conclusions**: This systematically validated framework combines adversarial training, attention-weighted discrimination, and adaptive loss scheduling to achieve improved accuracy, clinically relevant lesion differentiation, and computational efficiency suitable for routine clinical workflows.

**Trial Registration Number:** Tabriz University of Medical Sciences Research Ethics Committee (IR.TBZMED.REC.1402.902)

# 1. INTRODUCTION

Multiple Sclerosis (MS) is a chronic, inflammatory, demyelinating disease of the central nervous system affecting approximately 2.8 million people worldwide [1]. Diagnosis, monitoring, and treatment planning for MS heavily rely on accurate assessment of brain structural changes visible on Magnetic Resonance Imaging (MRI), particularly white matter hyperintensities (WMH) and ventricular alterations [2]. These imaging biomarkers are crucial for evaluating disease progression, treatment efficacy, and prognosis.

White matter hyperintensities appear as bright regions on T2-weighted and fluid-attenuated inversion recovery (FLAIR) MRI sequences, representing areas of demyelination, inflammation, and tissue damage [3]. Their spatial distribution, count, and volumetric progression correlate with clinical disability measures and cognitive impairment [4]. Furthermore, the burden and evolution patterns of these hyperintensities have been associated with both disease progression and therapeutic response [3]. Historical studies have established WMH as independent predictors of cognitive decline and functional outcomes across various neurological conditions [5]. Ventricular enlargement serves as an important indicator of brain atrophy, another hallmark of MS progression [6,7].

Current clinical practice relies heavily on qualitative visual assessment by radiologists who manually identify and characterize lesions across multiple MRI slices [2]. This process is time-consuming, labor-intensive, and subject to significant inter-observer variability, potentially leading to inconsistent diagnosis and treatment decisions [8]. The complex nature of MS pathology requires precise differentiation between pathological hyperintensities and normal appearing white matter signal variations, a distinction challenging even for experienced specialists [9].

Despite advances in neuroimaging analysis, several critical limitations persist in automated segmentation tools for neurodegenerative diseases. First, almost no current approach simultaneously addresses ventricle and WMH segmentation within a unified framework, despite their anatomical and pathophysiological relationship [10]. Umirzakova et al. [11] demonstrated that ventricle morphology and proximity significantly influence the interpretation of periventricular hyperintensities, yet most segmentation approaches treat these structures independently. This disconnected approach fails to capitalize on contextual information that could improve segmentation accuracy.

Second, current methods struggle to differentiate between normal hyperintensities resulting from cerebrospinal fluid (CSF) contamination and pathological lesions. This distinction is crucial for accurate disease quantification, as Griffanti et al. [12] and Dadar et al. [13] emphasized that misclassification of normal appearing white matter signal variations as lesions can lead to overestimation of disease burden. McKinley et al. [7] demonstrated that up to 30% of automatically detected hyperintensities in routine clinical scans may represent normal anatomical variants rather than pathology.

Third, the mismatch between research methodologies and clinical data characteristics presents a significant obstacle. Most state-of-the-art segmentation methods assume isotropic 3D volumes with high resolution in all dimensions [14-22]. However, routine clinical MRI protocols typically produce anisotropic data with significant slice thickness, creating a fundamental incompatibility. Dong et al. [23] highlighted how this anisotropy presents substantial challenges for conventional 3D segmentation approaches, which may either fail entirely or require computationally intensive preprocessing steps. Recent work by Tran et al. [2] further demonstrated that inconsistencies in imaging protocols and resolution contribute significantly to variability in WMH quantification, particularly affecting the detection of small lesions and periventricular boundaries.

The computational demands of 3D deep learning models present another practical barrier to clinical implementation. While these models have demonstrated impressive performance in research settings, Tran et al. [24] found that their deployment in typical clinical environments is constrained by hardware limitations and workflow integration challenges. This computational burden is particularly problematic in resource-limited healthcare settings where advanced computing infrastructure may not be available.

To address these limitations, we propose a novel 2D pix2pix deep learning framework specifically designed for the simultaneous segmentation of ventricles and white matter hyperintensities in anisotropic clinical MRI data. Our approach builds upon recent innovations in generative adversarial networks for medical image segmentation that have demonstrated improved handling of structural complexity and boundary precision in neuroimaging applications [25,26]. Building upon our previous work on pix2pix-based gray matter segmentation for WMH categorization [27], building on this foundation, the present work extends the pix2pix methodology to simultaneously segment multiple brain structures critical for MS evaluation, and introduces a systematic architectural ablation study to identify the optimal combination of adversarial training components for this multi-class task. Our method uniquely distinguishes between normal periventricular hyperintensities and pathological lesions, offering a comprehensive solution compatible with routine clinical workflows. The pix2pix architecture [28] serves as the foundation of our approach, modified to accommodate our multi-class segmentation task.

Unlike methods requiring resource-intensive 3D processing, our slice-based approach offers significant advantages for clinical implementation. As highlighted by Hossain et al. [29] and Hashemi et al. [30], 2D slice-based methods can achieve comparable accuracy to 3D approaches while dramatically reducing computational requirements, particularly important for handling anisotropic clinical scans. Our proposed model segments four distinct classes: background, ventricles, normal WMH (CSF-contaminated hyperintensities), and abnormal WMH (pathological lesions). This multi-class approach represents a significant advancement over existing tools that typically segment either ventricles or WMH independently.

A key strength of our work is the substantial dataset comprising 300 clinical MRI scans with comprehensive expert annotations of ventricles and both types of WMH. Our model incorporates FLAIR sequence as an input, enabling more accurate differentiation between periventricular CSF-contaminated hyperintensities and true pathological lesions [31,32]. Our method is designed with clinical utility as a primary goal, focusing on compatibility with standard clinical MRI protocols without requiring specialized acquisition parameters or time-consuming preprocessing. This practical approach significantly distinguishes our work from many research-oriented segmentation methods that perform well on research-grade datasets but fail to translate effectively to routine clinical data.

This article continues with detailed descriptions of our dataset and annotation protocol, the five-variant pix2pix segmentation architecture, training and evaluation strategy including 5-fold cross-validation, quantitative and qualitative results of both the ablation study and baseline comparisons, followed by a discussion of clinical implications and limitations.

## 2. MATERIALS & METHODS

## 2.1. Study Population

This study analyzed data from 300 MS patients imaged at Golgasht Medical Imaging Center, Tabriz, Iran. The cohort comprised 79 males (aged 18-57 years, mean = 35.8, SD = 9.3) and 221 females (aged 18-68 years, mean = 37.8, SD = 9.7). The study protocol received approval from the Tabriz University of Medical Sciences Research Ethics Committee (IR.TBZMED.REC.1402.902), and all participants provided written informed consent.

## 2.2. MRI Acquisition Protocol

All imaging was performed on a 1.5-Tesla TOSHIBA Vantage scanner (Canon Medical Systems, Japan). The T2-FLAIR images were acquired with the following parameters: TR = 10000 ms, TE = 100 ms, TI = 2500 ms, flip angle = 90°, FOV = 230 × 230 $mm^2$, acquisition matrix = [0, 256, 192, 0], resulting in non-isotropic data with voxel dimensions of (0.9, 0.9, 6) millimeters. This substantial difference between in-plane resolution and slice thickness informed our decision to implement a 2D rather than 3D segmentation approach.

## 2.3. Ground Truth Development

Manual segmentation was performed on all 300 MS patients' FLAIR images by a neuroradiologist with over 20 years of expertise in neuroimaging, supported by a trained medical imaging specialist. A custom Python-based GUI facilitated the annotation process, displaying T1-weighted, T2-weighted, and FLAIR sequences simultaneously to enable cross-modal verification. Four distinct classes were delineated: (1) background, (2) ventricles, (3) normal white matter hyperintensities (WMH), and (4) abnormal WMH.

### 2.3.1. Annotation Workflow

**Phase 1 - Primary Manual Annotation**: Initial delineation was performed slice-by-slice in the following order: (1) ventricles, identified by very low FLAIR intensity and anatomical location; (2) abnormal WMH (MS lesions), identified based on high FLAIR intensity combined with MS-characteristic features including periventricular or juxtacortical location, ovoid morphology perpendicular to ventricles, minimum diameter of 3 mm, and confirmation on T2-weighted images; and (3) normal WMH, defined as hyperintensities lacking MS-specific characteristics, typically appearing as thin periventricular rims or bilateral symmetric age-related changes.

**Phase 2 - Statistical Morphological Refinement**: Each structure underwent automated post-processing to improve consistency while preventing over- or miss-segmentation. For ventricles, morphological closing (3×3 structuring element) filled small gaps and smoothed boundaries, followed by connected component analysis to remove isolated regions (< 5 pixels). Intensity-based validation confirmed that mean intensities fell below the 25th percentile for ventricles and exceeded the 75th percentile for both WMH classes, with failing cases flagged for review. For WMH, conservative morphological opening (3×3 structuring element) removed artifacts, with substantial changes (> 10% area) triggering manual verification. These percentile-based thresholds ensured adaptability to inter-patient intensity variations.

**Phase 3 - Expert Consensus Review**: The senior neuroradiologist independently reviewed all cases with overlaid masks at adjustable transparency. Discrepancies were discussed and resolved through consensus, with systematic documentation of refinements to standardize criteria.

**Phase 4 - Final Quality Control**: Both experts conducted comprehensive review of consensus annotations, with flagged or unusual cases undergoing detailed re-examination and manual correction as needed.

### 2.3.2. Periventricular Normal WMH Definition

Normal WMH identification required special handling due to CSF contamination effects. The periventricular boundary was established by dilating ventricle masks using a disk-shaped structuring element with 5-pixel radius (approximately 4.5 mm), consistent with established CSF partial volume effect ranges (3-5 mm). High-intensity pixels within this boundary (exceeding the 75th percentile of brain tissue intensity) were identified, regions overlapping with abnormal WMH were excluded, and remaining candidates underwent manual refinement focusing on frontal and occipital horn vicinities. The expert verified that each region exhibited bilateral symmetry, smooth margins, and absence of MS-characteristic features.

This comprehensive protocol—combining systematic manual delineation, conservative statistical refinement, and multi-stage expert review—ensured accurate differentiation between normal hyperintensities (CSF contamination and age-related changes) and abnormal hyperintensities (true MS lesions). The ventricular system was coded in blue, abnormal WMH in red, and normal WMH in green, as shown in **Fig. 1**.

## 2.4. Public Dataset

To enhance the generalizability and robustness of our findings, we incorporated the publicly available MSSEG 2016 challenge dataset [33]. This dataset comprises 15 MS patients' MRI data acquired from 3 different imaging centers with standardized acquisition protocols. The original MSSEG 2016 dataset provides expert-annotated ground truth segmentations only for white matter hyperintensities (MS lesions). Since our methodology requires simultaneous segmentation of both ventricles and WMH, we supplemented this dataset by having our neuroradiologist expert manually segment the ventricular regions and normal WMH using the same protocol described in section 2.3. This comprehensive annotation enables training and evaluation of our proposed multi-class segmentation approach across diverse imaging conditions and patient populations.

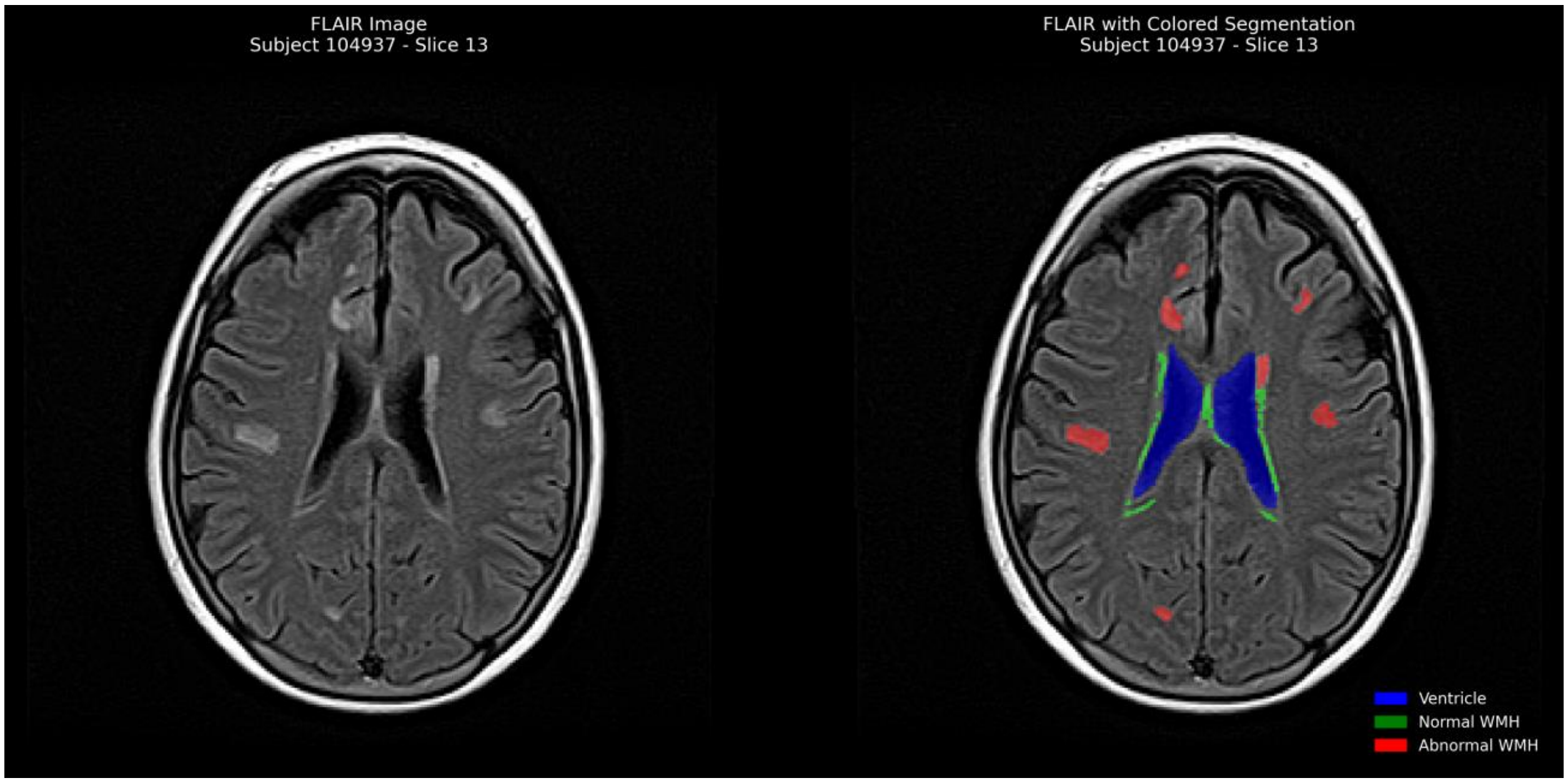

**Fig. 1** A sample image of MRI FLAIR data with corresponding manual segmentation and annotation. The left image shows an original FLAIR scan from an MS patient. The right image displays the tri-mask overlay with abnormal WMH (red), normal WMH (green), and ventricles (blue).

## 2.5. Pre-processing Pipeline

Our pre-processing pipeline forms a critical foundation for successful application of deep learning for brain MRI segmentation, as illustrated in **Fig. 2**. This pipeline standardizes input data, reduces noise, and prepares images for optimal model training and inference.

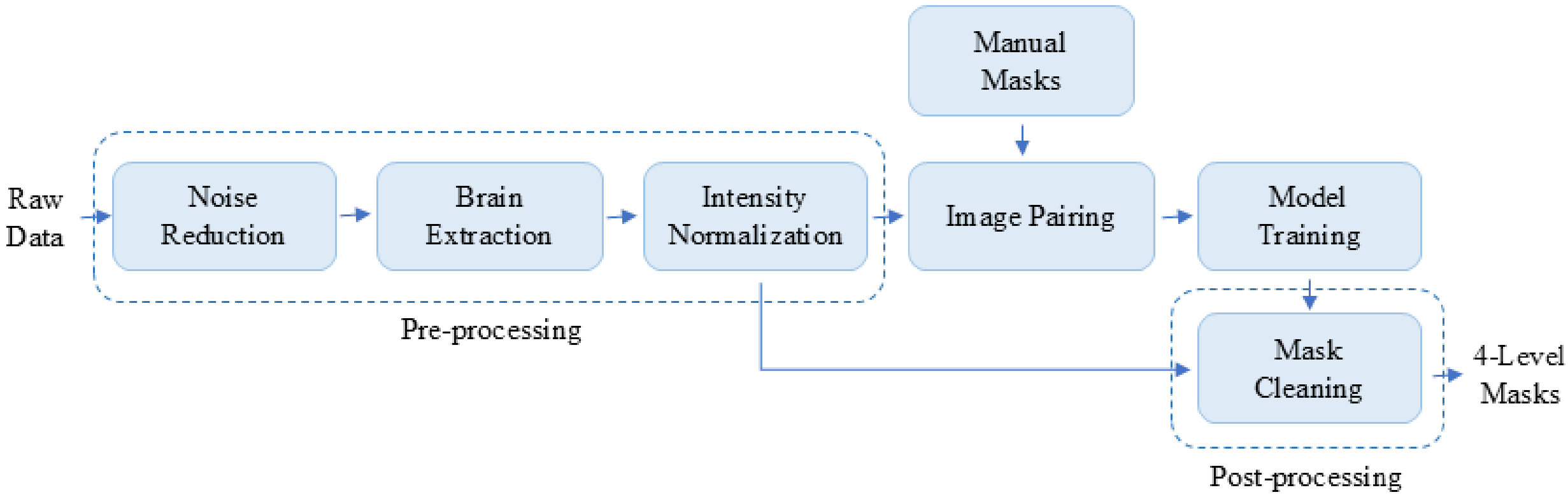

**Fig. 2** Block diagram of our proposed framework.

### 2.5.1. Noise Reduction

MR images inherently contain various noise artifacts that impede effective segmentation. We employ a two-stage noise reduction strategy that applies a median filter (kernel size = 3×3) to suppress salt-and-pepper noise while preserving edges, followed by selective application of a Gaussian filter ($\sigma$ = 1.0) to smooth remaining artifacts while retaining structural details. We deliberately apply minimal preprocessing, as our deep learning architecture is designed to handle more sophisticated noise patterns, including bias field inhomogeneity, directly during training following evidence that deep convolutional networks can learn to distinguish signal from noise given sufficient training examples [34].

### 2.5.2. Brain Extraction

Non-brain tissue was removed using the Brain Extraction Tool (BET) from FSL (FMRIB's Software Library) [35]. BET employs a deformable model approach that evolves a tessellated surface to fit the brain boundary, effectively separating brain tissue from skull, scalp, and other non-cerebral structures. The tool was applied to the FLAIR volume with default parameters, generating a binary brain mask that delineates the brain region while preserving internal structures including ventricles and white matter.

The resulting brain mask serves multiple purposes: eliminating non-brain tissues that could introduce artifacts, establishing a standardized region for intensity normalization (Section 2.5.3), and defining the region of interest for subsequent scaling operations. This preprocessing step reduces background noise and focuses computational resources on clinically relevant structures, improving segmentation accuracy and efficiency.

### 2.5.3. Intensity Normalization

We implement a slice-based intensity normalization technique addressing both inter- and intra-patient intensity variations. Unlike conventional global approaches, our method establishes parameters independently for each image slice. The minimum intensity value derives from average background intensity specific to each slice, while the maximum intensity value is determined by analyzing peripheral structures (scalp, skull, and surrounding tissues) obtained by subtracting the brain mask from the full image. This approach ensures intra-patient consistency since background and skull structures remain relatively constant across slices from the same patient, while providing inter-patient adaptability by accommodating anatomical variations. Recent work by Huang et al. [36] has demonstrated that such adaptive intensity normalization strategies significantly improve segmentation performance, particularly for detecting subtle intensity variations characteristic of MS lesions. Intensity values are scaled to the range [0,1] using Eq. 1, where normalization parameters are derived using our slice-specific methodology. Eventually, we apply z-score normalization on the scaled values to secure more robust normalized intensities.

$$I_{normalized} = \frac{I - I_{min}}{I_{max} - I_{min}} \quad (1)$$

### 2.5.4. Paired-image Generation

The final pre-processing step prepares input for conditional Generative Adversarial Network (cGAN) architecture. The pre-processed image and its ground truth mask are horizontally concatenated to create a paired 256×512 pixel composite image compatible with the pix2pix architecture [28]. This paired-image approach enables the model to learn the mapping between MR images and their segmentation masks within a unified spatial context.

## 2.6. Network Architecture and Implementation

Our method employs a cGAN architecture—specifically the pix2pix model—as the core component for brain segmentation. As illustrated in **Fig. 2**, this model bridges the pre-processing stages and post-processing refinements in our pipeline. The detailed network architecture and implementation strategy are described in the following and visually represented in **Fig. 3**.

### 2.6.1. Pix2Pix Framework

We leverage the pix2pix framework for its proven effectiveness in image-to-image translation while preserving structural coherence. The framework consists of two networks: a generator (modified U-Net) that translates MR images into segmentation masks and a discriminator (PatchGAN) that evaluates mask realism. As depicted in **Fig. 3**, the generator receives the pre-processed MR images and produces segmentation masks, while the discriminator simultaneously learns to distinguish between generated masks and ground truth annotations. This adversarial objective encourages the generator to produce segmentations that are not only voxel-accurate but also anatomically plausible, improving boundary coherence in structurally complex regions such as periventricular zones.

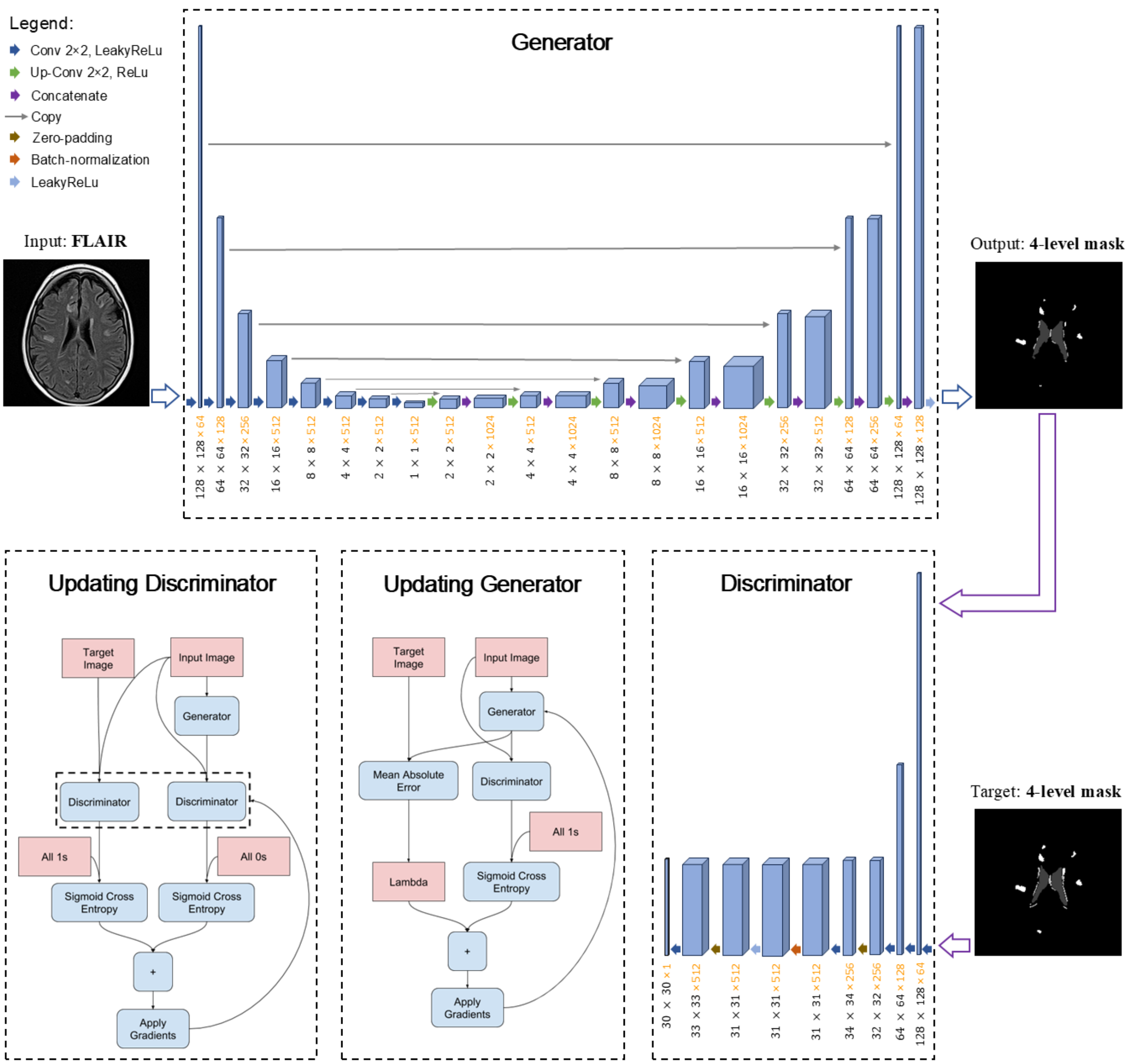

**Fig. 3** Pix2pix model architecture overview. This system is trained using paired images consisting of input FLAIR sequences and corresponding 4-level segmentation masks as targets. The updating blocks displayed along the bottom row illustrate the learning progression of the respective network components.

### 2.6.2. U-Net Backbone

The generator employs a U-Net architecture [37] with an encoder–decoder structure and skip connections, adapted for multi-class segmentation. The encoder comprises four downsampling blocks (convolutional layer, batch normalization, leaky ReLU activation), followed by a bottleneck layer, while the decoder includes four corresponding upsampling blocks (transposed convolutions, batch normalization, ReLU activation, selective dropout). Skip connections between encoder and decoder layers at equivalent spatial resolutions preserve fine-grained structural detail critical for accurate boundary delineation.

The PatchGAN discriminator evaluates the realism of predicted segmentation masks by classifying overlapping image patches rather than the full image, enabling focused assessment of local boundary quality. In our final architecture (V5, described in Section 2.6.3), the discriminator incorporates an attention mechanism that applies class-specific weighting to its outputs: foreground classes (ventricles, normal WMH, and abnormal WMH) receive twice the weight of background regions, explicitly counteracting the severe class imbalance inherent in medical segmentation tasks where foreground structures represent a small fraction of total image voxels.

### 2.6.3. Architectural Variants and Training Strategy

#### Architectural Variants

To systematically evaluate the contribution of each architectural component to segmentation performance, we implemented and compared six model variants under identical training conditions:

- V0 — Baseline U-Net: Standard U-Net trained with weighted cross-entropy (WCE) loss, without the adversarial training component. Serves as the non-GAN baseline.
- V1 — Baseline Pix2Pix with WCE: The pix2pix architecture adapted for multi-class segmentation, trained with WCE as the segmentation loss component.
- V2 — Baseline Pix2Pix with UFL: Replaces WCE with unified focal loss (UFL) to address class imbalance through dynamic down-weighting of well-classified examples.
- V3 — Pix2Pix + Attention Discriminator + WCE: Adds the attention mechanism to the PatchGAN discriminator (described in Section 2.6.2), retaining WCE as the segmentation loss.
- V4 — Pix2Pix + Adaptive Hybrid Loss: Replaces the fixed loss function with an adaptive hybrid combining WCE and UFL, without the attention discriminator.
- V5 — Pix2Pix + Attention Discriminator + Adaptive Hybrid Loss (Final Model): Integrates both the attention discriminator and the adaptive hybrid loss. This is our final proposed architecture.

**Loss Function Formulation**

The total loss for all pix2pix variants (V1–V5) combines an adversarial component and a segmentation component:

$$\mathcal{L}_{total} = \lambda_{GAN} \cdot \mathcal{L}_{GAN} + \lambda_{seg} \cdot \mathcal{L}_{seg}\ , \ where \ \ \lambda_{GAN} = 1 \ and \ \lambda_{seg} = 50. \quad (2)$$

where $\mathcal{L}_{GAN}$ is the adversarial loss (binary cross-entropy between real and generated mask assessments by the discriminator), and $\mathcal{L}_{seg}$ is the segmentation loss.

For V4 and V5, $\mathcal{L}_{seg}$ is the adaptive hybrid loss that smoothly transitions from WCE to UFL over the course of training:

$$\mathcal{L}_{seg} = (1 - \beta(t)) \cdot \mathcal{L}_{WCE} + \beta(t) \cdot \mathcal{L}_{UFL} \tag{3}$$

$$\beta(t) = \sigma(\frac{\left(\frac{t}{T}\right) - h}{k}) \tag{4}$$

where $\beta(t)$ is a sigmoid function of the current epoch $t$, total epochs $T$, and constants $k$ and $h$ controlling the transition speed and midpoint, respectively. This formulation enables stable early convergence through WCE while progressively emphasizing hard-example mining through focal loss in later epochs. For V0 and V1, $\mathcal{L}_{seg} = \mathcal{L}_{WCE}$; for V2, $\mathcal{L}_{seg} = \mathcal{L}_{UFL}$.

**Dataset Splits and Cross-Validation**

Patient-level stratification was performed separately for local and public datasets to prevent subject-level data leakage. Both datasets were combined under a unified split:

- Training: 210 patients (local) + 9 patients (public) → 4,650 images
- Validation: 30 patients (local) + 3 patients (public) → 750 images
- Test: 60 patients (local) + 3 patients (public) → 1,350 images

All individual patient slices remained within the same partition, ensuring that no patient contributed images to more than one split. Five-fold cross-validation was applied with patient-level stratification: the test split remained fixed across all folds, while validation assignments were rotated across non-test patients such that no two folds shared validation subjects. The best generator checkpoint per fold was selected based on lowest validation loss.

**Training Configuration**

All six variants were trained under identical conditions: batch size = 4, epochs = 60, Adam optimizer ($\beta_1$ = 0.9, $\beta_2$ = 0.999), initial learning rate = $2 \times 10^{-4}$ with scheduled reduction. Generator and discriminator were updated alternately following the standard GAN training procedure. Variant-specific hyperparameters (e.g., focal loss $\gamma$, hybrid transition constants k and h) are documented in the training scripts available in the repository (see Section 2.9).

#### 2.6.4. Post-processing

The raw network output undergoes post-processing to improve segmentation quality and anatomical plausibility. We apply a series of morphological operations to the predicted masks such as connected component analysis, hole filling, and edge smoothing. These minimal post-processing steps preserve the network's predictions while removing obvious artifacts, resulting in high-precision segmentation of target brain structures with anatomical plausibility.

## 2.7. Baseline Methods

To evaluate the performance of our proposed approach, we compare it against established methods for both ventricle and hyperintensity segmentation. Since no existing method simultaneously performs four-class segmentation (background, ventricles, normal WMH, abnormal WMH), we conducted separate comparisons for ventricle and hyperintensity segmentation.

#### 2.7.1. Ventricle Segmentation Baselines

SynthSeg: We employ the SynthSeg model, a deep learning-based approach trained on synthetic data for robust brain MRI segmentation [14]. Following the official implementation guidelines, we apply SynthSeg to our FLAIR images and extracted ventricle masks from the full brain parcellation output based on their corresponding label values.

Atlas Matching: We implement an atlas-based approach using the MNI152 standard space template [38]. The process involves registering each subject's FLAIR image to MNI152 space using FSL's linear registration tool (FLIRT), applying the ventricle atlas mask, performing basic morphological post-processing to refine boundaries, and finally transforming the resulting segmentation back to the subject's native space.

#### 2.7.2. Hyperintensity Segmentation Baselines

BIANCA (Brain Intensity AbNormality Classification Algorithm): We utilize this FSL-based supervised method for WMH segmentation [31]. Following the recommended protocol, BIANCA is trained on our manual segmentations using the same training dataset split as our proposed method to ensure fair comparison.

LST-LPA (Lesion Segmentation Tool - Lesion Prediction Algorithm): This unsupervised lesion segmentation algorithm is implemented under the SPM toolbox [39]. We apply LST-LPA directly to our FLAIR images without parameter modification, as recommended in the official documentation. The output binary masks are obtained in the subject's native space.

LST-LGA (Lesion Segmentation Tool - Lesion Growth Algorithm): This algorithm, also part of the SPM toolbox, requires both FLAIR and T1-weighted images [40]. We pre-register T1 images to the corresponding FLAIR space using FSL's FLIRT command and apply LST-LGA with default parameter settings. Binary output masks are generated in the subject's native space.

WMH-SynthSeg: This extension of the SynthSeg framework specifically addresses white matter hyperintensity segmentation [6]. Following the official guidelines, we apply WMH-SynthSeg to our FLAIR images and extract the WMH masks based on the designated label in the output.

### 2.7.3. Adaptation for Fair Comparison

For comparing our approach with baseline methods, several adaptations are necessary. Since our method distinguishes between normal and abnormal WMH while baseline methods produce only a single WMH class, we use abnormal WMH predictions for direct comparison with the baseline hyperintensity segmentation methods. Similarly, as our model simultaneously segments both ventricles and WMH, we extract the ventricle class predictions for comparison with ventricle-specific baselines. All evaluations are performed in the subject's native space to avoid registration-induced errors in the quantitative comparisons.

## 2.8. Evaluation Metrics and Analysis

To comprehensively assess the performance of our proposed method and compare it with the baseline methods, we employ multiple complementary evaluation metrics. These metrics are calculated separately for ventricle segmentation and hyperintensity segmentation.

### 2.8.1. Detection-based Metrics

Precision: Also known as positive predictive value, precision measures the proportion of correctly identified positive voxels among all voxels classified as positive:

$$Precision = \frac{TP}{TP+FP} \quad (5)$$

where TP represents true positives and FP represents false positives.

Recall: Also known as sensitivity or true positive rate, recall measures the proportion of actual positive voxels that were correctly identified:

$$Recall = \frac{TP}{TP+FN} \tag{6}$$

where FN represents false negatives.

### 2.8.2. Overlap Metrics

Dice Similarity Coefficient (DSC): The Dice coefficient quantifies the spatial overlap between two segmentations and is defined as:

$$DSC = \frac{2\,|A \cap B|}{|A|+|B|} \tag{7}$$

where $A$ represents the predicted segmentation and $B$ represents the ground truth segmentation. DSC ranges from 0 (no overlap) to 1 (perfect overlap).

Jaccard Index (JI): Also known as the Intersection over Union (IoU), the Jaccard index is calculated as:

$$JI = \frac{|A \cap B|}{|A \cup B|} \tag{8}$$

JI also ranges from 0 to 1 and is mathematically related to the Dice coefficient.

### 2.8.3. Boundary Metrics

Hausdorff Distance: While the traditional Hausdorff distance measures the maximum distance between the boundaries of two segmentations, it is highly sensitive to outliers. Therefore, we used the 95th percentile Hausdorff distance (HD95), which is more robust to small segmentation errors:

$$HD95(A, B) = \max\left(h_{95}(A, B),\ h_{95}(B, A)\right) \tag{9}$$

where $h_{95}(A, B)$ is the 95th percentile of the distances from points in boundary $A$ to their closest points in boundary $B$. HD95 is measured in millimeters, with lower values indicating better boundary agreement.

### 2.8.4. Performance Evaluation

To evaluate the performance of normal versus abnormal WMH classification, we compute a confusion matrix at the lesion level. This approach allows for assessment of how well the algorithm distinguishes between the two types of hyperintensities as distinct entities.

## 2.9. Development Environment and Technical Implementation

The deep learning framework was implemented using Python 3.9 with TensorFlow, OpenCV, and scikit-image for image preprocessing, and NumPy and Pandas for data handling and statistical analyses. All experiments ran on a workstation with an Intel Core i7-7700K CPU, 64 GB RAM, and NVIDIA RTX 3060 GPU. The codebase with documentation, preprocessing pipelines, model architectures, training scripts, and evaluation modules is available on GitHub (https://github.com/Mahdi-Bashiri/Sim-Vent-WMH-Seg) under an MIT license, including pre-trained model weights for immediate inference.

# 3. RESULTS

## 3.1. Model Training and Convergence

The cGAN architecture demonstrated efficient training characteristics and robust convergence behavior across all segmentation tasks. Training progression analysis revealed consistent improvement patterns for all three target structures (ventricles, normal WMH, and abnormal WMH).

### 3.1.1. Architectural Variant Comparison

The ablation study across the six architectural variants (V0–V5) reveals a clear and consistent performance progression, demonstrating the additive contribution of each component to the overall segmentation quality. Results are summarized in **Table 1** and training dynamics of fold 4 are visualized in **Fig. 4**, which presents loss convergence and validation metrics across all six variants with best-epoch selection marked for V1–V5.

The baseline U-Net (V0), trained without adversarial supervision, achieved the lowest mean Dice (0.714 ± 0.018), reflecting the inherent difficulty of optimizing boundary precision with pixel-wise loss alone on imbalanced multi-class data. Despite achieving the highest recall (0.937 ± 0.002), this came at the expense of substantially reduced precision (0.616 ± 0.020) and the worst HD95 (6.50 ± 0.46 mm), indicating extensive false-positive regions.

Introducing the adversarial training component (V1, baseline pix2pix with WCE) produced an immediate and substantial improvement: mean Dice rose to 0.823 ± 0.011, precision improved to 0.805 ± 0.005, and HD95 reduced to 5.31 ± 0.20 mm. This confirms that the discriminator's boundary-level feedback is the primary driver of segmentation quality improvement in this task, consistent with prior findings on GAN-based medical image segmentation [25,26].

**Table 1.** Comparative performance of six architectural variants for simultaneous ventricle and WMH segmentation evaluated on independent test set (n=60 patients) using 5-fold cross-validation. Metrics represent mean±SD across 5 folds. Bold indicates best performance per metric.

| Variant | Architecture | mean DSC | mean IoU | mean HD95 | mean Precision | mean Recall |
|---|---|---|---|---|---|---|
| V0 | Baseline U-Net | 0.714 ±0.018 | 0.601 ±0.020 | 6.50 ±0.46 | 0.616 ±0.020 | **0.937** ±0.002 |
| V1 | Baseline Pix2Pix with WCE | 0.823 ±0.011 | 0.721 ±0.014 | 5.31 ±0.20 | 0.805 ±0.005 | 0.848 ±0.010 |
| V2 | Baseline Pix2Pix with UFL | 0.817 ±0.010 | 0.714 ±0.012 | 5.50 ±0.35 | 0.791 ±0.021 | 0.850 ±0.019 |
| V3 | Pix2Pix + Attention Discriminator + WCE | 0.824 ±0.008 | 0.723 ±0.011 | 5.23 ±0.31 | 0.807 ±0.014 | 0.843 ±0.009 |
| V4 | Pix2Pix + Adaptive Hybrid Loss | 0.844 ±0.002 | 0.750 ±0.003 | **4.81** ±0.05 | 0.845 ±0.009 | 0.843 ±0.010 |
| V5 | Pix2Pix + Attention Discriminator + Adaptive Hybrid Loss | **0.852** ±0.004 | **0.760** ±0.006 | 4.87 ±0.13 | **0.856** ±0.006 | 0.850 ±0.006 |

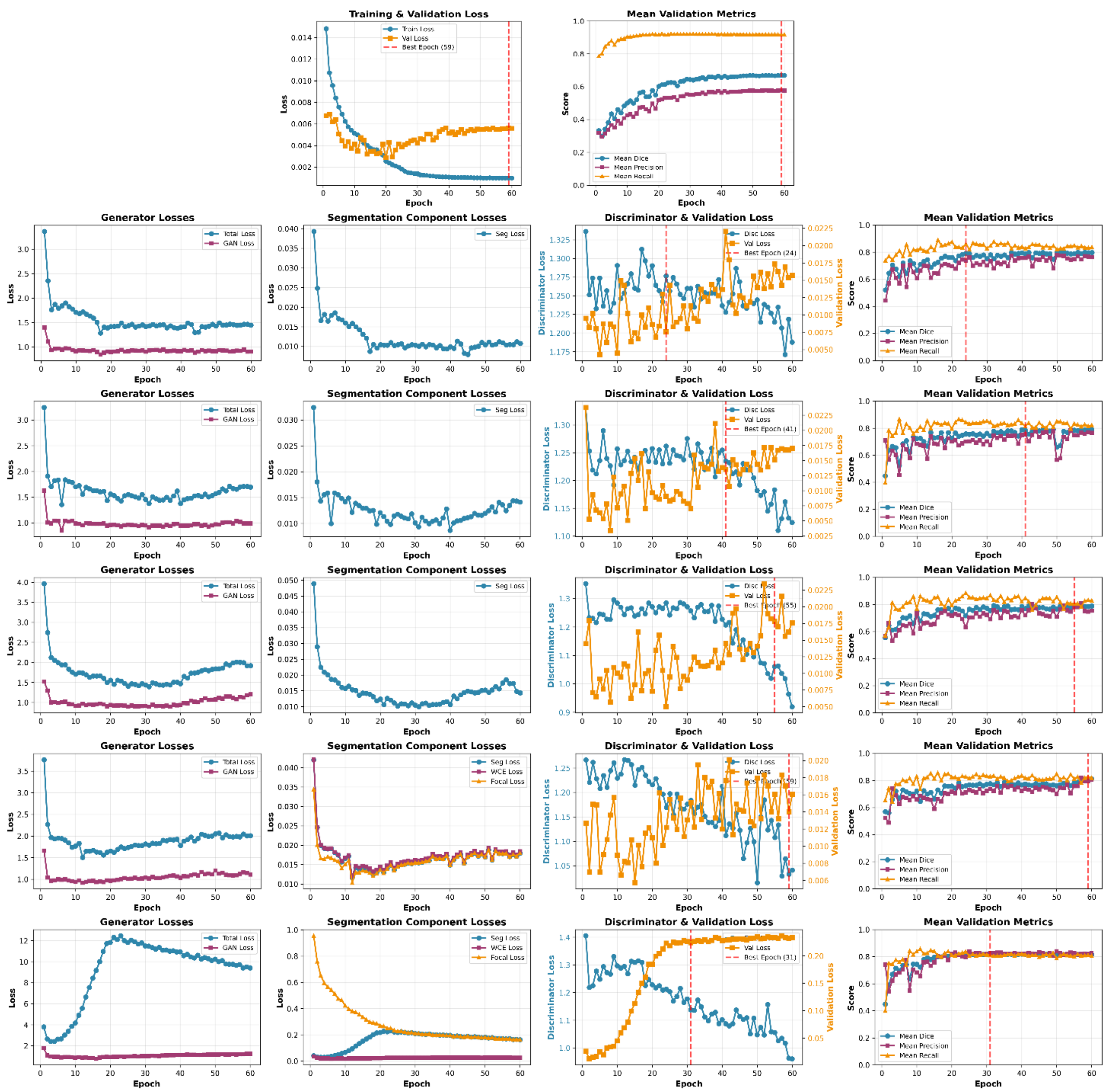

**Fig. 4**. Training dynamics and convergence behavior across six architectural variants (V0–V5) in fold 4. Top row (V0): Baseline U-Net without adversarial training shows only training/validation loss and mean validation metrics. Rows 2–6 (V1–V5): Pix2pix-based variants display four panels per row: (left to right) generator losses, segmentation component losses, discriminator & validation losses, and mean validation metrics. The vertical dashed line in the rightmost two panels indicates the epoch selected for final model evaluation based on optimal validation performance across all segmentation classes. Note the progressive improvement in validation metrics and convergence stability from V1 (baseline pix2pix) through V5 (attention discriminator + adaptive hybrid loss).

Substituting the loss function from WCE to unified focal loss (V2) yielded comparable performance to V1 (Dice: 0.817 ± 0.010), suggesting that neither loss function alone is sufficient to fully address the class imbalance present across the four segmentation targets.

Adding the attention mechanism to the discriminator (V3, WCE) produced a marginal but consistent improvement over V1 (Dice: 0.824 ± 0.008, HD95: 5.23 ± 0.31 mm), confirming that directing the discriminator's attention toward foreground classes yields a meaningful signal for handling imbalanced structures such as normal WMH.

The most significant single-component gain was achieved by the adaptive hybrid loss (V4), which raised mean Dice to 0.844 ± 0.002 and HD95 to 4.81 ± 0.05 mm — the best boundary accuracy of any variant. The smooth transition from WCE to UFL across training epochs enables stable early convergence while progressively emphasizing hard examples in later epochs, a strategy particularly effective for small and heterogeneous WMH lesions.

The final architecture V5, combining both the attention discriminator and adaptive hybrid loss, achieved the best overall performance: mean Dice 0.852 ± 0.004, mean IoU 0.760 ± 0.006, mean Precision 0.856 ± 0.006, and mean Recall 0.850 ± 0.006, with HD95 of 4.87 ± 0.13 mm. V5 is selected as our proposed method for all subsequent comparisons.

### 3.1.2. Per-Class Performance of the Final Model

Per-class analysis of V5 on the independent test set reveals consistent performance across all three foreground classes, as detailed in **Table 2**.

**Table 2.** Per-class segmentation performance of the final model (V5) on the independent test set, evaluated across 5-fold cross-validation. Metrics represent mean ± SD.

| Class | mean DSC | mean IoU | mean HD95 | mean Precision | mean Recall |
|---|---|---|---|---|---|
| Ventricles | 0.907 ±0.002 | 0.830 ±0.004 | 3 ±0.51 | 0.916 ±0.005 | 0.899 ±0.007 |
| Normal WMH | 0.677 ±0.007 | 0.512 ±0.008 | 4.87 ±0.24 | 0.660 ±0.019 | 0.696 ±0.010 |
| Abnormal WMH | 0.825 ±0.009 | 0.703 ±0.013 | 4.51 ±0.32 | 0.849 ±0.013 | 0.804 ±0.021 |
| Ventricles & Abnormal WMH | 0.852 ±0.004 | 0.760 ±0.006 | 4.87 ±0.13 | 0.856 ±0.006 | 0.850 ±0.006 |

Ventricles achieved the highest segmentation performance (Dice: 0.907 ± 0.002, HD95: 3.00 ± 0.51 mm), consistent with their relatively well-defined boundaries and characteristic low FLAIR intensity. Abnormal WMH segmentation demonstrated strong performance (Dice: 0.825 ± 0.009, HD95: 4.51 ± 0.32 mm), reflecting the model's effective learning of MS lesion distribution patterns despite their heterogeneous size and shape. Normal WMH — the most challenging class due to their thin periventricular distribution and proximity to both CSF and pathological lesions — achieved a Dice coefficient of 0.677 ± 0.007 and HD95 of 4.87 ± 0.24 mm, demonstrating that the spatial context learned from simultaneous ventricular segmentation meaningfully supports this most ambiguous class.

## 3.2. Segmentation Performance: Qualitative Analysis

### 3.2.1. Visual Representation of Segmentation Results

**Fig. 5** presents representative axial FLAIR slices with overlaid segmentation results, using a consistent color scheme of true positives (green), false positives (red), and false negatives (yellow). V5 demonstrated precise boundary delineation for ventricles and enhanced detection of both small and large WMH lesions across both local and public dataset cases. Error visualization revealed that false positives predominantly occurred at peripheral boundaries, while false negatives were primarily located in regions with lower contrast or partial volume effects.

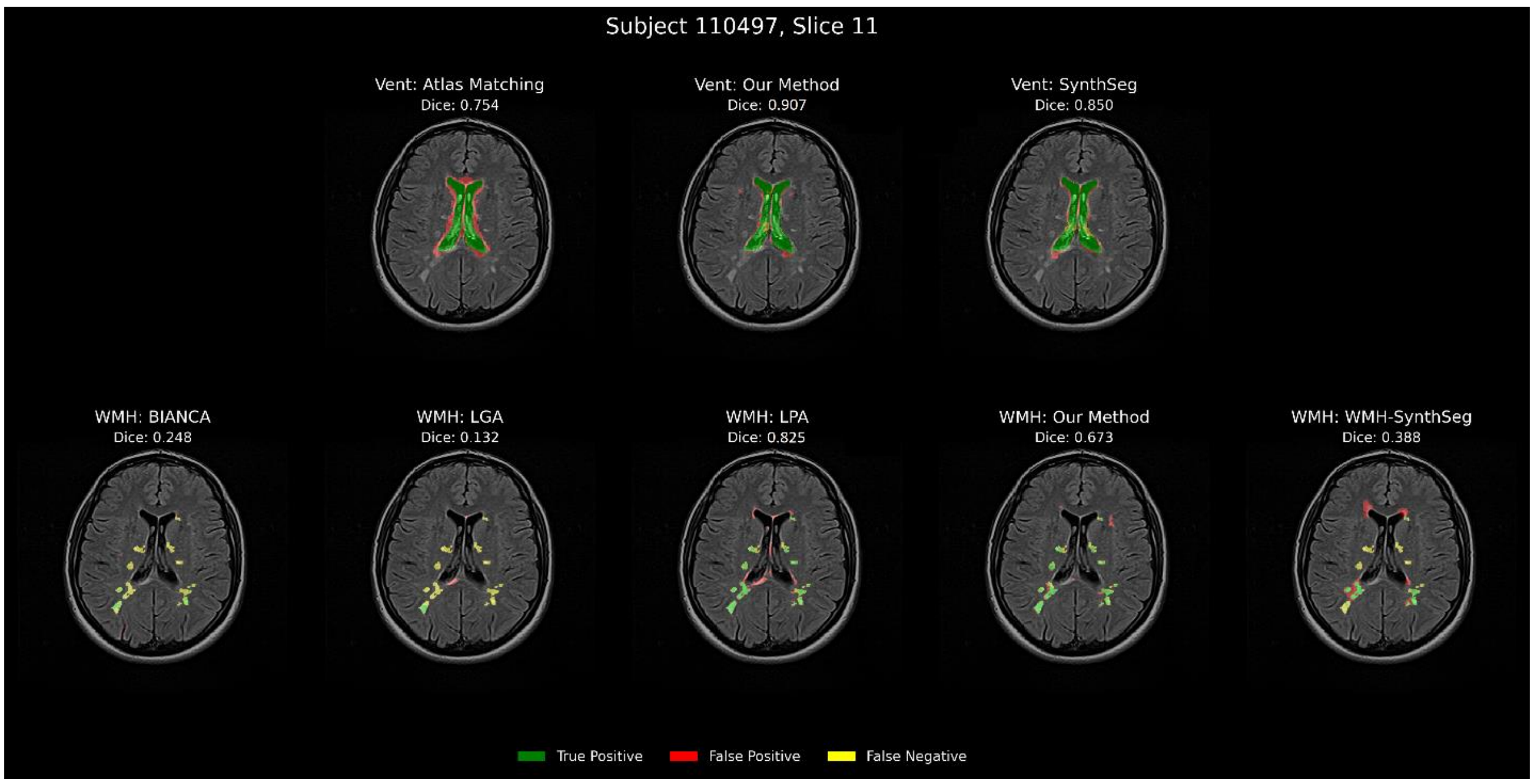

**Fig. 5** A sample FLAIR image with predicted masks. The first and second rows show the results of ventricle and WMH segmentation, respectively. Color-coding indicates true positives (green), false positives (red), and false negatives (yellow) for visual comparison.

### 3.2.2. Comparative Visual Analysis

Ventricle Segmentation Comparison: The three ventricle segmentation methods exhibited distinct characteristics: Atlas Matching produced substantial false positives at ventricular boundaries and adjacent sulcal spaces, demonstrating limitations when applied to MS patients with anatomical variations. SynthSeg showed more conservative segmentation with moderate false negatives, particularly in the inferior portions of lateral ventricles and third ventricle. Our method achieved balanced performance with minimal false positives and negatives. Segmentation boundaries closely followed ventricular margins even in challenging regions such as ventricular horns.

WMH Segmentation Comparison: The five WMH segmentation methods demonstrated varying capabilities: LGA showed significant under-segmentation with extensive false negatives, particularly for smaller and less hyperintense lesions. BIANCA exhibited substantial false negatives with limited detection of periventricular lesions characteristic of MS pathology. LPA demonstrated improved lesion detection but increased false positives, frequently misclassifying normal periventricular hyperintensities as pathological. WMH-SynthSeg showed moderate false negatives for smaller and challenging lesions and substantial false positives in periventricular regions. Our method achieved superior performance with extensive true positives encompassing both large and small lesions while maintaining minimal false positives and negatives. Notably, our approach successfully differentiated between normal periventricular hyperintensities and pathological MS lesions.

## 3.3. Segmentation Performance: Quantitative Analysis

### 3.3.1. Ventricle Segmentation Metrics

Comprehensive Performance Metrics: Quantitative evaluation of ventricle segmentation across both datasets demonstrates the consistent superiority of our proposed method (V5) (**Table 3**). V5 achieved the highest Dice coefficient (0.907 ± 0.002) and recall (0.899 ± 0.007), with competitive precision (0.916 ± 0.005) and the best boundary accuracy (HD95: 3.00 ± 0.51 mm). On the public dataset, V5 demonstrated strong and consistent performance, confirming that the combined-dataset training strategy with patient-level stratification yields robust generalization across different acquisition protocols and scanner types.

**Table 3.** Comprehensive performance metrics for ventricle segmentation methods across local and public datasets (mean ± standard deviation).

| Method | Precision | Recall | Dice | Jaccard | Hausdorff95 |
|---|---|---|---|---|---|
| Atlas Matching | 0.748 ± 0.114 | 0.771 ± 0.122 | 0.737 ± 0.113 | 0.590 ± 0.126 | 7.6 ± 6.32 |
| SynthSeg | 0.851 ± 0.105 | 0.785 ± 0.058 | 0.810 ± 0.08 | 0.692 ± 0.112 | 6.1 ± 6.57 |
| Our Method (V5) | 0.916 ± 0.005 | 0.899 ± 0.007 | 0.907 ± 0.002 | 0.830 ± 0.004 | 3.0 ± 0.51 |

SynthSeg showed competitive performance but demonstrated substantially higher variance (HD95: 6.1 ± 6.57 mm), indicating sensitivity to the imaging characteristics of specific cases. Atlas Matching performed consistently below both deep learning methods on boundary accuracy. Our V5 method maintained the lowest standard deviations across all metrics, reflecting the stability conferred by 5-fold cross-validation and the larger combined training cohort.

### 3.3.2. White Matter Hyperintensity Segmentation Metrics

Comprehensive Performance Metrics: Quantitative evaluation of WMH (abnormal class) segmentation confirms the superior and consistent performance of our proposed method V5 (**Table 4**). V5 achieved the highest Dice (0.825 ± 0.009), precision (0.849 ± 0.013), and recall (0.804 ± 0.021), alongside the best boundary accuracy (HD95: 4.51 ± 0.32 mm). These results represent a substantial improvement over all baseline methods. Performance differences between methods are most pronounced, where our method was trained with the majority of the data.

### 3.3.3. Normal vs. Abnormal Hyperintensity

Confusion Matrix Analysis: The confusion matrix for V5's differentiation between normal and abnormal WMH (**Fig. 6**) demonstrates strong performance with notable asymmetry. The model achieved a true positive rate of 79.85% for abnormal hyperintensities and a true negative rate of 93.65% for normal hyperintensities. The false positive and false negative rates were 6.35% and 20.15% respectively, revealing a conservative bias toward classifying ambiguous WMH as normal rather than abnormal. This error pattern, where the model is more likely to underestimate pathological lesions (false negatives) than to overestimate them (false positives), reflects the inherent difficulty in distinguishing true MS lesions from CSF-contaminated periventricular signal.

Classification Metrics: V5 demonstrated strong discriminative performance between the two WMH classes with an overall accuracy of 84.28%. Recall (sensitivity) for abnormal WMH was 79.85%, with specificity of 93.65% for normal WMH and precision of 96.38% for pathological classification. The F1-score of 0.8734 for this binary differentiation task confirms meaningful clinical discrimination between CSF-contaminated periventricular signal and true MS lesions. Cohen's kappa of 0.6707 indicates substantial agreement beyond chance, despite the challenging nature of this classification boundary.

**Table 4.** Comprehensive performance metrics for WMH segmentation methods across local and public datasets (mean ± standard deviation).

| Method | Precision | Recall | Dice | Jaccard | Hausdorff95 |
|---|---|---|---|---|---|
| BIANCA | 0.405 ± 0.239 | 0.201 ± 0.175 | 0.269 ± 0.143 | 0.155 ± 0.095 | 33.68 ± 17.59 |
| LGA | 0.748 ± 0.138 | 0.247 ± 0.155 | 0.371 ± 0.159 | 0.228 ± 0.145 | 35.43 ± 16.62 |
| LPA | 0.637 ± 0.294 | 0.450 ± 0.237 | 0.527 ± 0.25 | 0.358 ± 0.21 | 27.64 ± 24.63 |
| WMH-SynthSeg | 0.604 ± 0.141 | 0.391 ± 0.127 | 0.475 ± 0.142 | 0.311 ± 0.12 | 34.99 ± 24.75 |
| Our Method (V5) | 0.849 ± 0.013 | 0.804 ± 0.021 | 0.825 ± 0.009 | 0.703 ± 0.013 | 4.51 ± 0.32 |

**Fig. 6** Confusion matrices of the final model (V5). The matrix shows the classification of abnormal WMH vs. normal WMH according to the segmentation of all WMH.

Error Analysis: The elevated false negative rate (20.15%) — substantially exceeding the false positive rate (6.35%) — reflects the inherent ambiguity at the boundary between normal and pathological periventricular hyperintensities, particularly in cases with confluent WMH where the distinction between CSF contamination effects and true lesion margins is not definitive even for expert annotators. While the high precision (96.38%) indicates that when the model predicts abnormal WMH it is highly reliable, the moderate recall (79.85%) suggests approximately one-fifth of true abnormal lesions are misclassified as normal. This conservative classification pattern, while avoiding false alarms, may lead to underestimation of lesion burden in clinical quantification and warrants consideration in longitudinal monitoring applications.

## 3.4. Computational Efficiency

### 3.4.1. Processing Time and Resource Analysis

Our comprehensive analysis of computational efficiency across all evaluated methods revealed substantial differences in processing time and resource requirements, as detailed in **Table 5**. Our proposed deep learning approach demonstrated exceptional efficiency, requiring approximately 4 seconds for end-to-end processing per patient case (2.4 seconds for pre-processing and 1.6 seconds for model inference and post-processing), dramatically outperforming all baseline methods.

**Table 5.** Computational Performance Comparison of Segmentation Methods.

| Method | Training | Inference | CPU | RAM | GPU |
|---|---|---|---|---|---|
| BIANCA | 58 sec | 11 sec | ~20% | ~1 GB | N/A |
| LST-LGA | N/A | 147 sec | ~60% | ~2 GB | N/A |
| LST-LPA | N/A | 72 sec | ~40% | ~2 GB | N/A |
| WMH-SynthSeg | N/A | 78 sec | 100% | ~5 GB | 100% |
| Atlas Matching | N/A | 115 sec | ~40% | ~2 GB | N/A |
| SynthSeg | N/A | 124 sec | 100 % | ~5 GB | 100% |
| Our Method | 257 sec* | <4 sec† | 15% | ~1 GB | 80% |

* Time per training epoch. Total training time: 261 minutes (60 epochs).

† Includes 2.4 sec for pre-processing and 1.6 sec for model inference and post-processing.

The detailed breakdown of execution times revealed marked differences across all methods. BIANCA achieved moderate performance with an 11-second processing time but demonstrated limited segmentation accuracy. The traditional atlas-based approaches (LST-LGA and Atlas Matching) exhibited the longest processing times at 147 and 115 seconds per case, respectively, making them less suitable for time-sensitive clinical applications. The deep learning-based SynthSeg and WMH-SynthSeg methods showed intermediate performance (124 and 78 seconds) but required full CPU utilization and considerable memory resources (~5 GB RAM and 100% GPU usage).

In contrast, V5's end-to-end inference time of approximately 4 seconds per case demonstrates that the architectural enhancements (attention discriminator, adaptive hybrid loss) introduce no meaningful inference overhead. This efficiency stems directly from the 2D slice-based approach and the streamlined pix2pix generator, which processes each FLAIR slice in approximately 1.6 seconds including post-processing, with the remaining ~2.4 seconds allocated to preprocessing. The training overhead of 5-fold cross-validation (total ~22 hours) is a one-time cost borne offline, not reflected in clinical deployment time.

# 4. DISCUSSION

This study introduces a novel deep learning approach for the simultaneous segmentation of ventricles and white matter hyperintensities in multiple sclerosis, with the unique capability to differentiate between normal and pathological hyperintensities. Our comprehensive evaluation across local and public datasets demonstrates that the proposed pix2pix-based framework surpasses existing methods in both segmentation accuracy and computational efficiency while revealing important insights about cross-dataset generalizability.

## 4.1. Technical Innovations and Performance Advantages

The superior performance of our final architecture (V5) can be attributed to three complementary technical contributions, each quantifiably validated through the ablation study.

Adversarial training as the primary performance driver. The most substantial single improvement arose from introducing adversarial supervision: the transition from baseline U-Net (V0, Dice: 0.714) to baseline pix2pix (V1, Dice: 0.823) represents a gain of 0.109 Dice points and a reduction in HD95 from 6.50 mm to 5.31 mm. This finding aligns with prior work demonstrating that discriminator feedback encourages sharper and more anatomically plausible boundaries beyond what pixel-wise loss can achieve alone [25,26]. The improvement was particularly pronounced in precision (0.616 → 0.805), confirming that adversarial training effectively suppresses false-positive detections — a clinically critical property for avoiding overestimation of lesion burden.

Attention-weighted discriminator for class imbalance. Adding the attention mechanism to the discriminator (V3 vs. V1) produced a consistent, if modest, improvement (Dice: 0.824 vs. 0.823, HD95: 5.23 vs. 5.31 mm). While the aggregate gain appears small, the mechanism targets the most underrepresented structures — particularly normal WMH, which occupies the smallest fraction of image voxels — by directing the discriminator's evaluative signal toward foreground regions through a 2:1 class-specific weighting. The contribution of this mechanism becomes fully realized in combination with the adaptive hybrid loss in V5, suggesting a synergistic interaction between boundary-level attention and hard-example emphasis.

Adaptive hybrid loss for progressive learning. The most impactful single architectural modification was the adaptive hybrid loss (V4 vs. V1: Dice 0.844 vs. 0.823, HD95: 4.81 vs. 5.31 mm). The smooth sigmoid-controlled transition from WCE to UFL across training epochs allows the model to benefit from the stable gradient signal of cross-entropy during early learning while progressively shifting emphasis toward hard examples — small WMH lesions and ambiguous periventricular boundaries — in later epochs. This two-phase learning strategy is particularly suited to the multi-class imbalance in our task, where the relative difficulty of each class changes as the model matures. The final combined architecture (V5) achieved mean Dice of 0.852 and HD95 of 4.87 mm, representing a 19.3% improvement in Dice and a 25.1% reduction in HD95 over the non-adversarial baseline (V0).

Simultaneous segmentation with contextual coupling. Our approach uniquely addresses the joint segmentation of ventricles and WMH within a unified framework, leveraging the anatomical and pathophysiological relationship between these structures. By simultaneously segmenting ventricles, normal WMH, and abnormal WMH, the model inherently learns their spatial co-occurrence patterns. This contextual coupling is particularly important for normal WMH identification: the periventricular boundary established by the ventricular segmentation provides spatial prior information that constrains the normal WMH class, reducing false-positive pathological detections in CSF-contaminated zones.

Differentiation of normal from pathological hyperintensities. The ability to distinguish between normal periventricular hyperintensities and pathological MS lesions (F1-score: 0.8734, overall accuracy: 84.28%) represents a clinically significant advancement over methods treating all hyperintensities as uniform. As Griffanti et al. [12] and Dadar et al. [13] demonstrated, misclassification of CSF-contaminated signal as pathological lesions leads to systematic overestimation of disease burden. Our method demonstrates high precision (96.38%) when classifying abnormal WMH, ensuring that predicted pathological lesions are highly reliable. However, the model exhibits a conservative bias with a false negative rate of 20.15% compared to a false positive rate of 6.35%, indicating that approximately one-fifth of true abnormal lesions are classified as normal. While this conservative approach avoids overestimation of lesion burden — a clinically preferable error mode given the implications for treatment decisions — it suggests the model prioritizes specificity over sensitivity in ambiguous periventricular regions. Cohen's kappa of 0.6707 indicates substantial agreement beyond chance, confirming meaningful clinical discrimination despite the inherent difficulty of this classification boundary where even expert annotators face ambiguity.

Compatibility with clinical anisotropic data. Our 2D slice-based approach directly addresses the fundamental incompatibility between most state-of-the-art 3D segmentation methods and routine clinical MRI protocols. The acquisition parameters of our dataset ($0.9 \times 0.9 \times 6.0$ mm$^3$) create a 6-fold anisotropy that invalidates the isotropic assumptions underlying 3D convolutional networks [23]. Rather than treating this as a limitation to be overcome through preprocessing, our architecture embraces it — processing each FLAIR slice at native in-plane resolution without spatial rescaling, preserving true image characteristics while enabling real-time inference.

## 4.2. Computational Efficiency and Clinical Implementation

The computational efficiency of V5 matches that of the original single-model implementation, confirming that the architectural enhancements introduce no inference overhead. With an end-to-end processing time of approximately 4 seconds per case, V5 is up to 36 times faster than traditional atlas-based methods (147 seconds for LST-LGA) and 18–31 times faster than other deep learning alternatives (SynthSeg: 124 seconds; WMH-SynthSeg: 78 seconds). This dramatic reduction, coupled with modest resource requirements (15% CPU utilization, ~1 GB RAM, 80% GPU usage), enables near real-time analysis that could be available during the same clinical session.

This efficiency stems from the 2D slice-based approach optimized for anisotropic clinical data, the inherently lightweight pix2pix generator architecture, and the removal of computationally expensive preprocessing steps such as spatial rescaling and image registration. The training overhead of 5-fold cross-validation (~22 hours total) is a one-time offline cost that does not affect deployment speed.

## 4.3. Clinical Implications and Validation

The combined-dataset training strategy and 5-fold cross-validation protocol provide a statistically rigorous foundation for clinical translation claims. By training and validating on data from multiple imaging centers simultaneously — rather than fine-tuning post hoc — the framework's performance on unseen patients reflects genuine generalizability rather than dataset-specific memorization. This distinction is practically important: clinicians deploying AI tools in new centers need confidence in cross-site performance without the burden of site-specific retraining.

## 4.4. Limitations and Future Directions

The differentiation between normal and abnormal hyperintensities remains challenging in cases with confluent periventricular lesions, where boundaries become inherently ambiguous. Our confusion matrix analysis revealed a higher false negative rate, suggesting challenges in distinguishing pathological lesions from normal hyperintensities in subtle cases.

While our combined training cohort encompasses two independent datasets (300 local patients; 15 MSSEG2016 patients across three imaging centers), the local dataset originates from a single 1.5-Tesla scanner with a standardized acquisition protocol. This homogeneity may limit the framework's robustness to the full variability of clinical MRI hardware encountered in multi-site deployment. The public dataset partially mitigates this through inclusion of different scanner types and field strengths, but broader multi-site validation remains an important step toward regulatory-grade clinical adoption. Additionally, the 2D processing approach, while optimal for the anisotropic data used here, limits exploitation of inter-slice spatial relationships that may benefit lesion detection in higher-resolution acquisitions.

Future work should priorities expanded validation across multi-site, multi-scanner datasets with diverse acquisition parameters, including 3-Tesla acquisitions, to assess the generalizability ceiling of the 2D approach. Integration of longitudinal information — tracking lesion evolution across serial examinations — would extend the framework's utility for disease progression monitoring and treatment response assessment. The framework's modular architecture is well-suited to integration within automated volumetric reporting pipelines, potentially enabling real-time quantitative MS assessment within existing radiology workflows. Development of user-friendly clinical interfaces and prospective validation studies would constitute the next stages toward regulatory consideration and routine deployment.

# 5. CONCLUSION

This study presents a systematically validated deep learning framework for simultaneous segmentation of ventricles and white matter hyperintensities in multiple sclerosis, with unique capability to differentiate normal periventricular hyperintensities from pathological MS lesions. Through structured ablation of five architectural variants under 5-fold cross-validation, we identified optimal integration of adversarial training, attention-weighted discrimination, and adaptive hybrid loss. The final architecture (V5) achieved mean Dice 0.852±0.004, with per-class performance of 0.907±0.002 (ventricles), 0.825±0.009 (abnormal WMH), and 0.677±0.007 (normal WMH)—representing 19.3% improvement over non-adversarial baseline. Combined training across local and public datasets with patient-level stratification strengthens generalizability. The 2D slice-based design enables 4-second inference on anisotropic clinical data—up to 36× faster than baseline approaches—with minimal hardware requirements. By combining rigorous architectural validation, clinically relevant differentiation, and computational efficiency, this work advances quantitative neuroimaging tools deployable within routine MS clinical workflows.

## Acknowledgements

We acknowledge Eng. Mehrdad Rahbarpour, Eng. Azad Ahmadi, and Dr. Aydin Ghalehasadi, colleagues at the Golgasht Medical Imaging Center for their assistance in acquiring data.

The authors acknowledge the use of Claude Sonnet 4 (Anthropic, accessed October 2025) for assistance in manuscript preparation, including writing support, and formatting assistance. All AI-generated content was carefully reviewed, edited, and validated by the authors to ensure accuracy and appropriateness for the scientific context.

## Declaration of Conflicting Interest

The authors declare no conflict of interest.

## Funding

This research received no specific grant from any funding agency in the public, commercial, or not-for-profit sectors.

# Ethics Approval Statement

Ethical approval was established by the Tabriz University of Medical Sciences Research Ethics Committee (IR.TBZMED.REC.1402.902). Also, the written approval letter was obtained from all patients.

# Data and Code Availability Statement

The code and sample test data that support the findings of this study are publicly available at https://github.com/Mahdi-Bashiri/Sim-Vent-WMH-Seg. The complete dataset used in this study is not publicly available due to privacy considerations and institutional policies. However, access to additional data may be granted to qualified researchers upon reasonable request directed to the corresponding author. Any inquiries regarding data access, implementation details, or further collaboration should be addressed to the corresponding author.

# Authors' contributions

All authors contributed equally to this work. M.B.B., M.S, and A.S.B. collectively conceived the study, developed the computational framework, performed the analysis, and drafted the manuscript. All authors participated equally in study design, data collection, validation, and critical manuscript revision. All authors read and approved the final manuscript.